\documentclass[10pt,twocolumn,aps,floatfix,citeautoscript,superscriptaddress,prl]{revtex4-2}
\usepackage[english]{babel}

\usepackage{graphicx}
\usepackage[]{subcaption}
\usepackage{natbib}
\usepackage{physics}

\usepackage{xcolor}
\usepackage{amsmath, amssymb}
\usepackage{multirow}
\usepackage{xcolor}
\usepackage{pgfplots} 
\usepackage{ulem}
\usepackage{listings}
\usepackage{amsfonts}
\usepackage{comment}
\usepackage[colorlinks=true, allcolors=blue]{hyperref}

\pgfplotsset{compat=1.18}

\definecolor{codegreen}{rgb}{0,0.6,0}
\definecolor{codegray}{rgb}{0.5,0.5,0.5}
\definecolor{codepurple}{rgb}{0.58,0,0.82}
\definecolor{backcolour}{rgb}{0.95,0.95,0.92}

\newif\ifcomments
\commentsfalse  

\newcommand{\qea}{\alpha^{BO}}

\graphicspath{{./}{figures/}}

\begin{document}

\title{Floquet Many-Body Cages}
\date\today

\author{Tom Ben-Ami}
\affiliation{Max-Planck-Institut f\"{u}r Physik komplexer Systeme, N\"{o}thnitzer Stra\ss e 38, Dresden 01187, Germany}
\affiliation{Theoretical Physics III, Center for Electronic Correlations and Magnetism, Institute of Physics, University of Augsburg, D-86135 Augsburg, Germany}

\author{Roderich Moessner}
\affiliation{Max-Planck-Institut f\"{u}r Physik komplexer Systeme, N\"{o}thnitzer Stra\ss e 38, Dresden 01187, Germany}

\author{Markus Heyl}
\affiliation{Theoretical Physics III, Center for Electronic Correlations and Magnetism, Institute of Physics, University of Augsburg, D-86135 Augsburg, Germany}
\affiliation{Centre for Advanced Analytics and Predictive Sciences (CAAPS), University of Augsburg, Universitätsstr. 12a, 86159 Augsburg, Germany}

\begin{abstract}
Many-body cages have very recently emerged as a general route for nonergodic behaviour in quantum matter.
Here, we show that new types of many-body cages can be engineered in Floquet circuits with the potential to realize novel nonequilibrium quantum states.
For that purpose, we first identify an explicit, general construction of Floquet circuits capable of hosting many-body cages.
We then present a generic strategy to engineer and structure Floquet many-body cages.
We demonstrate the developed scheme for the quantum hard disk model as a generic constrained model system, realizable for instance in Rydberg atom arrays.
We construct Floquet circuits yielding Floquet many-body cages with topological properties and $\pi$-quasienergy modes, implying `time crystalline' spatiotemporal order.
Our results can be directly extended to general quantum circuits, thus providing a new tool to engineer nonequilibrium behaviour in driven systems. 
%
\end{abstract}

\maketitle

\phantomsection \label{sec:intro}
\textit{Introduction~---~}
Over the last two decades the study of non-equilibrium quantum matter has witnessed remarkable progress, driven by capabilities of modern quantum simulators to probe dynamics far from thermal equilibrium at an unprecedented level~\cite{rydbergs, ultracoldatoms, trappedions, squbits}.
A particular focus has been on mechanisms leading to nonergodic behaviour, such as many-body or disorder-free localisation and Hilbert-space fragmentation~\cite{basko2006metal, smith2017disorder, karpov2021disorder, brighi2023hilbert, sala2020ergodicity}.
Such mechanisms are not only key for understanding fundamental questions of thermalisation and the foundations of statistical mechanics. They also enable the realization of novel phases and types of long-range order by violating the principle of equal a priori probabilities, thereby evading equilibrium constraints such as those imposed by the Mermin--Wagner theorem~\cite{mermin1966absence}. Examples include localisation-protected quantum order~\cite{grover2014quantum, schulz2019stark, van2019bloch, rudner2013anomalous}, and spatiotemporal forms of order~\citep{khemani2016phase, khemani2019brief, 2018RPPh...81a6401S, yao2018time, RevModPhys.95.031001}.

Recently, the concept of many-body cages (MBCs) has emerged as a novel mechanism for nonergodic behaviour in quantum matter~\cite{ben2025many, tan2025interference, nicolau2025fragmentation, jonay2025localized}. Here, local constraints in the Hamiltonian give rise to eigenstates localised on a subgraph of the many-body state graph (`Fock graph') by means of quantum interference.
These lead to flat bands in the many-body spectrum, localising information and suppressing transport.
Initial works further suggested the emergence of a novel nonequilibrium phase in the form of eigenspectrum order -- a many-body caged spin glass, with intricate, fractal-like properties~\cite{ben2025many}.

Against this backdrop, it is natural to ask: can MBCs exist under external driving, and, if so, can they be  constructed with controllable properties and functionalities?

In this work, we show that many-body cages can be stabilised and engineered within Floquet quantum circuits.
First, we provide a general and explicit construction of chiral Floquet circuits capable of hosting many-body cages. 
We introduce palindromic drives, Floquet circuits that satisfy a discrete time-reversal symmetry within each driving period, as a prototypical example of chiral drives which can generate MBCs.
Second, we show that Floquet engineering on the state graph enables MBCs to realize directly in Fock space topological motifs known from single-particle physics (such as SSH chains), independent of real-space geometry.
Third, we demonstrate our scheme using the quantum hard-disk model as a minimal yet generic system for constrained quantum matter realizable in state-of-the-art Rydberg atom arrays.
For this model, we construct a Floquet protocol that generates $\pi$-quasienergy modes. 

This elevates the many-body caged spin glass to a spatiotemporally ordered many-body caged discrete time crystal. Crucially, this order is enabled by local constraints and quantum interference rather than quenched disorder as in conventional many-body localised discrete time crystal.
Beyond this specific example, our results are directly applicable to a broad class of quantum circuits and models with local constraints, thereby enabling the realisation of diverse structured many-body cages with single-particle topological properties. 

\begin{figure*}[!htb]
    \centering
    \includegraphics[width=1.0\linewidth]{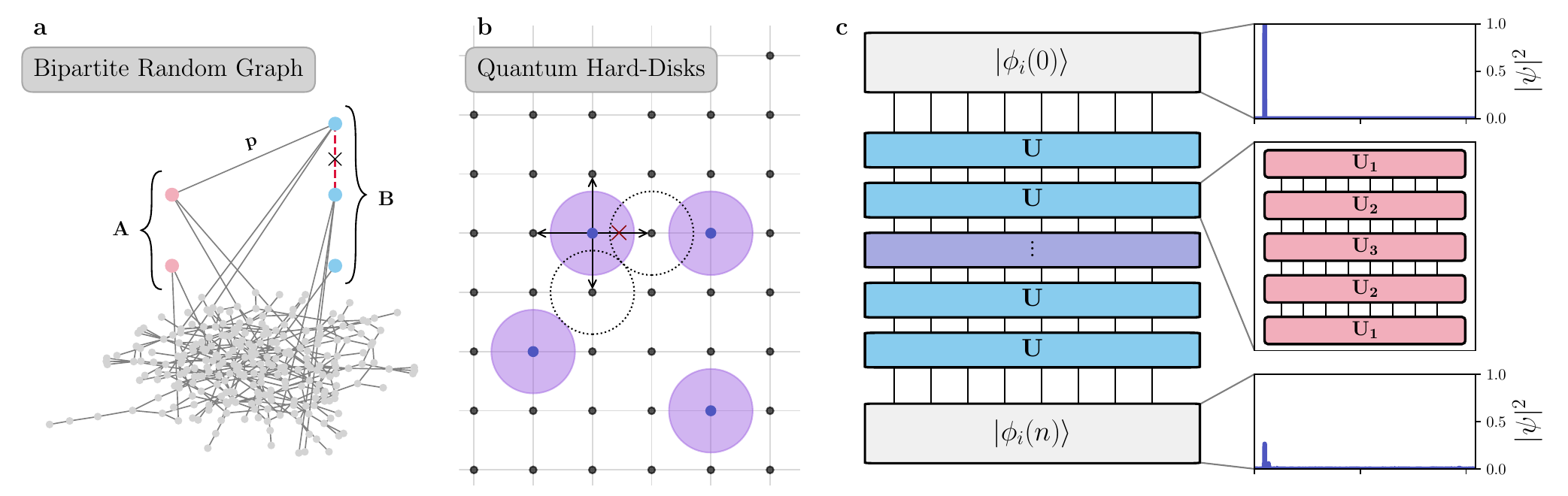}
    {\phantomsubcaption\label{sfig:IBRG}}
    {\phantomsubcaption\label{sfig:QHD}}
    {\phantomsubcaption\label{sfig:dipole}}
    \caption{\textbf{Floquet Driving of Caged Systems.} \textbf{a} - Schematic of an imbalanced bipartite random graph (IBRG). Edges between different sublattices $(A\leftrightarrow B)$ occur with probability $p$, while edges within the same sublattice are forbidden. \textbf{b} -  Schematic of the QHD model on the square lattice. \textbf{c} - Prescription of the Floquet palindromic drive. An initial and final caged state of the QHD model after $10^4$ periods, with $L=6, N=15$. The final state includes clear memory of the initial state.}
    \label{fig:Circuit}
\end{figure*}

\textit{Cages \& Models~---~} \phantomsection \label{sec:CM}
Many-body cages are eigenstates of a Hamiltonian $H$ localised on a subgraph of the many-body state graph, a representation of $H$, of a quantum system. Here, each node of the graph corresponds to a many-body basis state of the chosen basis (e.g. Fock basis), while the edges connect pairs of nodes $i$ and $j$ whenever $\mel{i}{H}{j}\neq0$. 

In general, MBCs can arise from various mechanisms. Here, we focus on two of those --- bipartite imbalance~\cite{jonay2025localized} and tree grafting~\cite{ben2025many, golinelli2003statistics, bauer2001random}.
The latter occurs in sparse state graphs, where compact localised states (CLS) emerge on weakly connected motifs, primarily dangling trees, but also structures such as grafted rings.
These MBCs depend only on their local neighbourhood in the state graph, where
local motifs support destructive interference.
The former, bipartite imbalance is rooted in a global symmetry --- chiral symmetry, where the state graph can be partitioned into two disjoint sublattices ${\mathcal{F}}_{1,2}$ (e.g., Fig.~\ref{sfig:IBRG}). When there is an imbalance with $N_{1} \not= N_2$ ($N_{1,2}$ the number of nodes on the two sublattices), the Hamiltonian spectrum possesses at least $|N_{1}-N_{2}|$ zero-energy modes.
Further details on both mechanisms are provided in the End Matter.

Although conceptually distinct, these mechanisms are not mutually exclusive, and both play a role in the examples discussed in this work.
Both mechanisms generate MBCs whose localised eigenstates display nonergodicity and memory effects.
Their localisation is induced by the graph structure itself, a feature central also for our targeted Floquet generalisation.

MBCs are particularly prominent in constrained quantum matter, where local constraints result in sparsely connected graphs. A paradigmatic model is the quantum hard-disk model (QHD)~\cite{ben2025many, naik2024quantum}:
\begin{equation} \label{eq:QHD}
	H = J\sum_{\left<i, j\right>} P_i \left( a_i^\dagger a_j + a_j^\dagger a_i \right) P_j,
\end{equation}
defined by hard-core particles on a 2D square lattice with excluded-volume interactions (Fig.~\ref{sfig:QHD}): configurations with two particles occupying neighbouring lattice sites are forbidden. Here, $a_i, a_j^{\dag}$ are hard-core bosonic annihilation/creation operators respectively, and $P_i$ is the projection operator enforcing the hard-disk constraint, 
$P_i = \prod_{j:\, |r_i - r_j| = 1} (1 - n_j)$, with $n_j \equiv a^\dagger_j a_j$.
Due to the bipartiteness of the real-space lattice, the state graphs of the QHD, and of similar kinetically constrained models, are also bipartite. 
This model is naturally realisable in state-of-the-art Rydberg atom arrays, where Rydberg blockade enforces the excluded volume constraint. 

The many-body state graphs of constrained models such as the QHD typically exhibit nonuniform connectivity and sparsely connected regions.
These qualitative features are shared by random graphs, which can thus serve as tunable and tractable effective models~\cite{ben2025many}.
Here, we focus on imbalanced bipartite random graphs (IBRGs) for systems exhibiting both bipartite structure and sublattice imbalance. We construct IBRGs by partitioning nodes into two distinct sets (sublattices), and adding edges only between nodes of distinct sets, with probability $p = \alpha/N$, where N is the total number of nodes, and $\alpha$ is the desired average connectivity. 

Systems with chiral symmetry have a block off-diagonal structure~\cite{altland1999field} in the sublattice resolved basis, which will be key for the Floquet circuits considered here,
\begin{equation} \label{eq:chiralH}
	\hat{H}^c = \mqty[ 0 & \hat{h}_{1\rightarrow2} \\ \hat{h}_{2\rightarrow1} & 0],
\end{equation}
where $1, 2$ denote the two sublattices. When systems additionally exhibit imbalance ($N_1 \neq N_2$), $\hat{h}$ is not a square matrix. This block structure appears naturally in both the QHD and IBRG models.

\phantomsection \label{sec:floq}
\textit{Chiral Floquet quantum circuits~---~}
In the following, we introduce a class of Floquet drives which host many-body cages. 
We consider Floquet circuits defined by a periodic unitary evolution $U(n) = U^n$, where $U = U_1 U_2 \cdots U_M$ is the single driving period unitary composed of $M$ layers. Each layer $U_j = \exp(-i H_j \tau_j)$ corresponds to the time evolution under a Hamiltonian $H_j$ over time $\tau_j$. From here on, we assume all Hamiltonians $H_j$ independently exhibit identical chiral symmetry. The effective dynamics over one period is generated by the Floquet Hamiltonian $H_F$ defined via $U = \exp(-i H_F\tau)$ where $\tau=\sum_j \tau_j$ denotes the period of the drive.

Since $H_F$ and $U$ share the same eigenbasis, constructing drives in which $H_F$ exhibits the structural properties that give rise to MBCs naturally extends their relevance to Floquet systems.  
We find that under chiral-symmetry-preserving drives, MBCs appear in $H_F$ through both bipartite imbalance and tree grafting.

To construct such chiral-symmetry preserving drives, we consider the Baker–Campbell–Hausdorff (BCH) expansion and obtain $H_F$ from a series of nested commutators of $H_j$'s.
Crucially, commutators of chiral-symmetric Hamiltonians exhibit characteristic structure: the commutator $[H_j, H_k]$ is block diagonal (chiral symmetry is broken).
However, the nested commutator $[H_j, [H_k, H_l]]$ of three chiral-symmetric Hamiltonians returns back to the block off-diagonal form (chiral symmetry is restored).
Thus, only nested commutators with an \textit{even} number of commutators (i.e., an odd number of Hamiltonian generators) preserve chiral symmetry. 
%
For that purpose, we introduce \textit{palindromic Floquet drives} of the form
\begin{equation}\label{eq:PalindromicDrive}
	U = U_1 U_2 \cdots U_{M} \cdots U_2 U_1,
\end{equation}
where the drive layers $U_j = \exp(-i H_j)$ are applied in time-reversed order about a central midpoint. This time-symmetric structure ensures that all odd-order commutators in the BCH expansion cancel exactly, leaving only even-order nested commutators that preserve chiral symmetry.
Consequently, the effective Hamiltonian $H_F$ inherits chiral symmetry structurally similar to that of the constituent chiral Hamiltonians. 

\begin{figure}[!htb]
    \centering
    \includegraphics[width=1.0\linewidth]{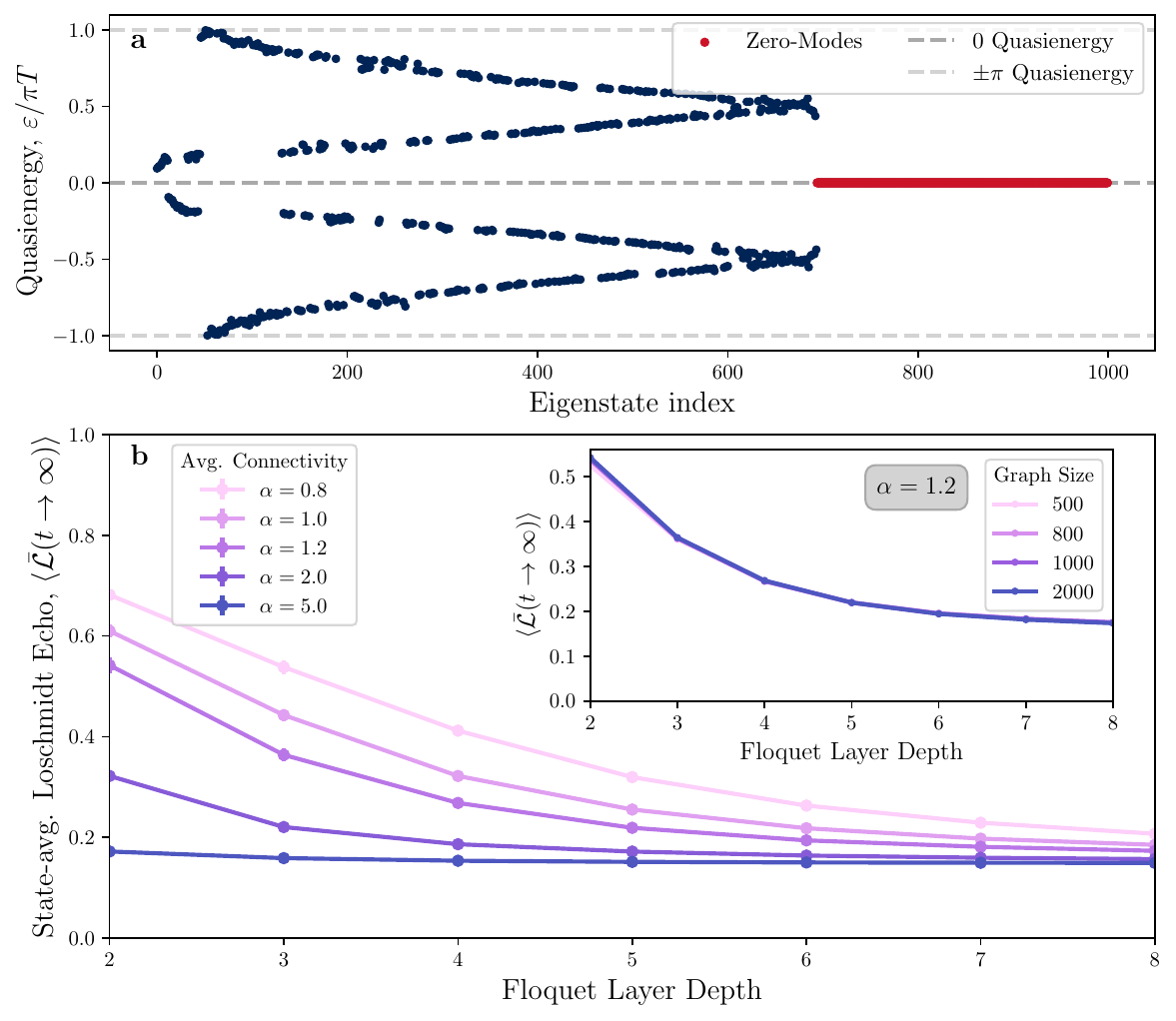}
    {\phantomsubcaption\label{sfig:RG_Spec}}
    {\phantomsubcaption\label{sfig:RG_LE}}
    \caption{\textbf{Long-time memory in driven IBRGs.} \textbf{a} Quasienergy specturm of the driven IBRG, $N=1000$. \textbf{b} Long time memory of the IBRG under increasing palindromic Floquet drive depth for difference values of average connectivity $(\alpha)$, with $N=2000$. As the depth increases, the effective $\alpha$ increases, and only memory arising from imbalance-induced cages remain. The inset shows effect of the graph size, for a single value of $\alpha$.}
    \label{fig:Floquet_Memory}
\end{figure}

\phantomsection \label{sec:obs}
\textit{Observables~---~} 
We detect and characterise Floquet MBCs using the Loschmidt echo as a generic, system-independent observable. 
For an initial state $\ket{\phi_i(0)}$ (here chosen as a basis state), the Loschmidt echo is  defined as the return probability after time evolution:
\begin{equation}
	\mathcal{L}_i(t) = \abs{\braket{\phi_i(t)}{\phi_i(0)}}^2.
\end{equation}
The Loschmidt echo quantifies the memory of an initial state and thereby provides a direct probe of nonergodicity.
In systems hosting MBCs with degenerate flat bands, the Loschmidt echo can display many-body Rabi oscillations at frequencies set by the flat-band energies of the cages --- a decisive signature of MBCs.
Loschmidt echoes have become increasingly accessible in quantum simulator experiments recently~\citep{karch2025probing, lunkin2026evidence}.

To assess general nonergodicity, we define a state-averaged Loschmidt echo, $\overline{\mathcal{L}}(t)$, as the uniform average over a complete set of initial product states. Nonergodic systems with persistent memory exhibit $1 \geq \bar{\mathcal{L}}(t\to\infty) > 0$, whereas $\bar{\mathcal{L}}(t\to\infty) \to 0$ signals ergodic behaviour. 

While $\mathcal{L}(t)$ probes initial-state memory retention of the time-evolved state and is a system-independent observable --- defined identically for any quantum system --- it is a global observable.
To complement this, for the QHD model, we also consider a local autocorrelation function (see End Matter) and discuss the similarities with the dynamics observed for the Loschmidt echo.

\phantomsection \label{sec:mbc_rg}
\textit{Floquet many-body cages in imbalanced bipartite random graphs~---~}
We demonstrate these considerations first for palindromic drives of the IBRGs.
Each layer of the Floquet protocol is generated from an independent IBRG with average connectivity $\alpha$; the palindromic symmetry then stitches these layers into a single symmetric circuit.

The quasienergy spectrum of such a Floquet drive, shown in Fig.~\ref{sfig:RG_Spec}, reveals a flat band at zero quasienergy.
Furthermore, the system exhibits long-time memory and therefore nonergodic behaviour, as seen from 
$\overline{\mathcal{L}}(t)$, plotted in Fig.~\ref{sfig:RG_LE} for different connectivities $\alpha$.
As the Floquet depth $M$ increases, memory is reduced due to fragmentation effects vanishing, but memory from cages persists. The inset of Fig.~\ref{sfig:RG_LE} shows that memory persists across a broad range of graph sizes, suggesting a nonvanishing value even in the thermodynamic limit and therefore that Floquet MBCs make up a finite fraction of Hilbert space.

\phantomsection \label{sec:mbc_hd}
\textit{Floquet many-body cages for quantum hard disks~---~}
We next turn to the experimentally relevant QHD model. We implement a horizontal-vertical (HV) drive with $H_1=H_V$ and $H_2=H_H$. Here, $H_V, H_H$ describe hard-core bosonic hopping along the vertical ($V$) and horizontal ($H$) directions of the square lattice with durations $\tau_V$ and $\tau_H$, respectively. For the palindromic drive we choose $U_{HV} = U_V U_H U_V$ with $U_H=\exp(-iH_H \tau_H)$ and $U_V=\exp(-i H_V \tau_V/2)$.
In the Supplementary Material we further show an adaptation of the Anomalous Floquet Anderson Insulator drive displaying similar properties.

\begin{figure}[!bht]
    \centering
    \includegraphics[width=1.0\columnwidth]{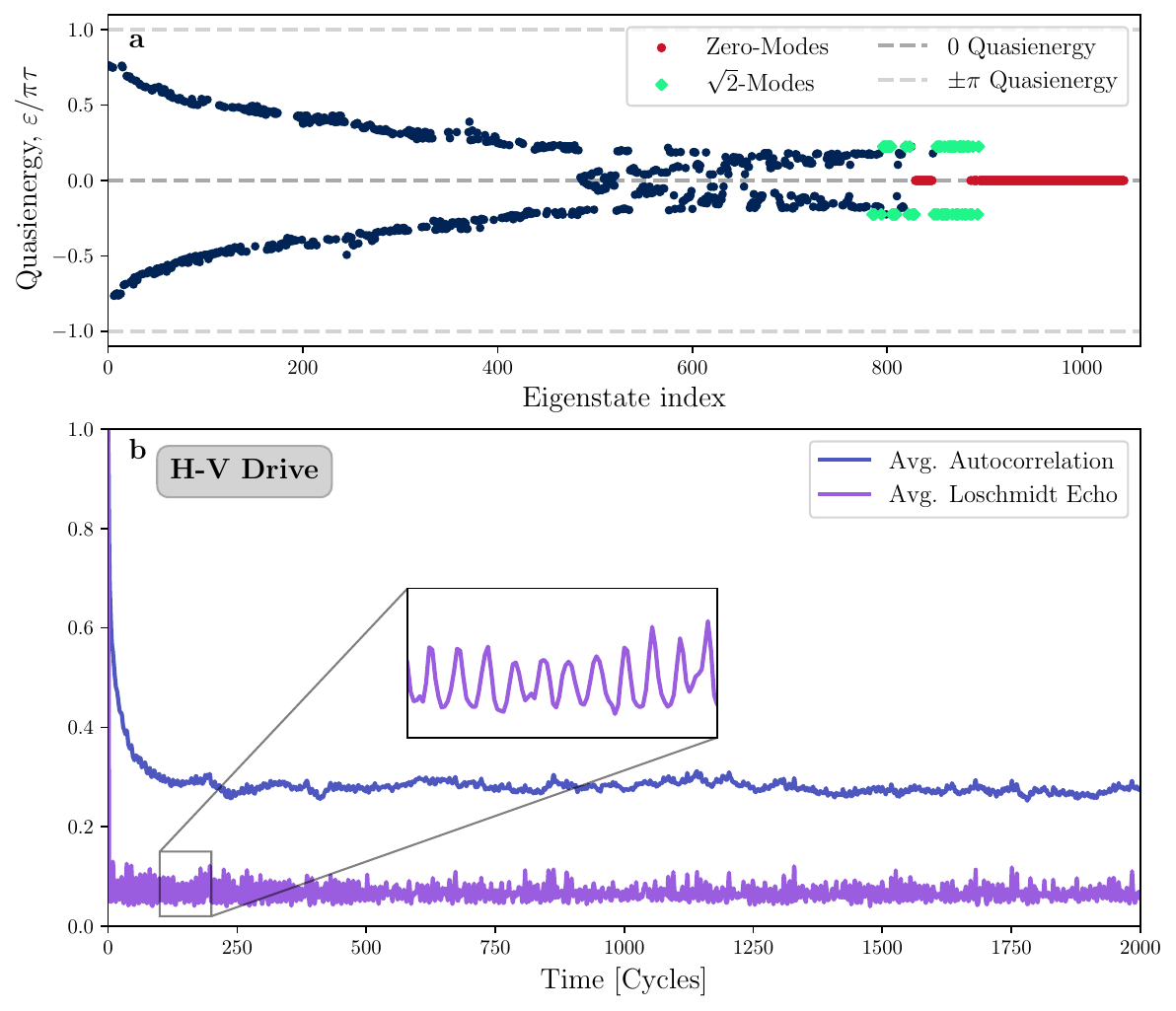}
    {\phantomsubcaption\label{sfig:HD_Spec}}
    {\phantomsubcaption\label{sfig:HD_LE}}
    \caption{\textbf{Floquet QHD model.} \textbf{a} Quasienergy spectrum of the QHD model under the HD drive, with $\tau_1 = \tau_2$. \textbf{b} The long-time memory of the QHD model is shown for the HV drive. Persistent oscillations, which arise from additional bands beyond $E=0$, can also be observed.
    }
    \label{fig:Floquet_QHD}
\end{figure}

As before, the palindromic drive implies zero-quasienergy MBCs through the chiral symmetry of $H_F$, see Fig.~\ref{sfig:HD_Spec} for the quasienergy spectrum for $\tau_V=\tau_H$. Note that this specific QHD configuration has no bipartite imbalance. 
Importantly, one can also identify a secondary band which corresponds to the originally static $E/J = \sqrt{2}$ band.
Thus, \textit{tree grafting} plays a dominant role in MBCs in this system, which will be key for the engineering introduced below.

The state-averaged Loschmidt echo $\overline{\mathcal{L}}(t)$ and autocorrelation, displayed in Fig.~\ref{sfig:HD_LE}, provide direct dynamical evidence for memory of initial states. Moreover, persistent oscillations at frequencies matching the quasienergy band separation confirm the presence of many-body Rabi oscillations -- a distinctive dynamical feature of MBCs.

\phantomsection \label{sec:eng}
\textit{Floquet engineering of many-body cages~---~}
Having established that palindromic drives preserve chiral symmetry and generate Floquet MBCs, we now demonstrate how to engineer these cages with novel spectral and topological properties. The key insight follows from the leading-order BCH expansion of the Floquet Hamiltonian for the HV drive protocol:	$H_F \tau \approx H_V \tau_V + H_H \tau_H + \mathcal{O}(\tau^3)$, where $\tau = \tau_V + \tau_H$ is the total period duration. By tuning the drive times $\tau_V$ and $\tau_H$, we effectively modulate the edge weights on the many-body state graph. 
Since the local state graph structure, and hence the MBC structure, emerges directly from the underlying constraints, this provides a powerful route to engineering multiple MBCs with tailored properties.

\begin{figure}[!hbt]
    \centering
    \includegraphics[width=1.0\columnwidth]{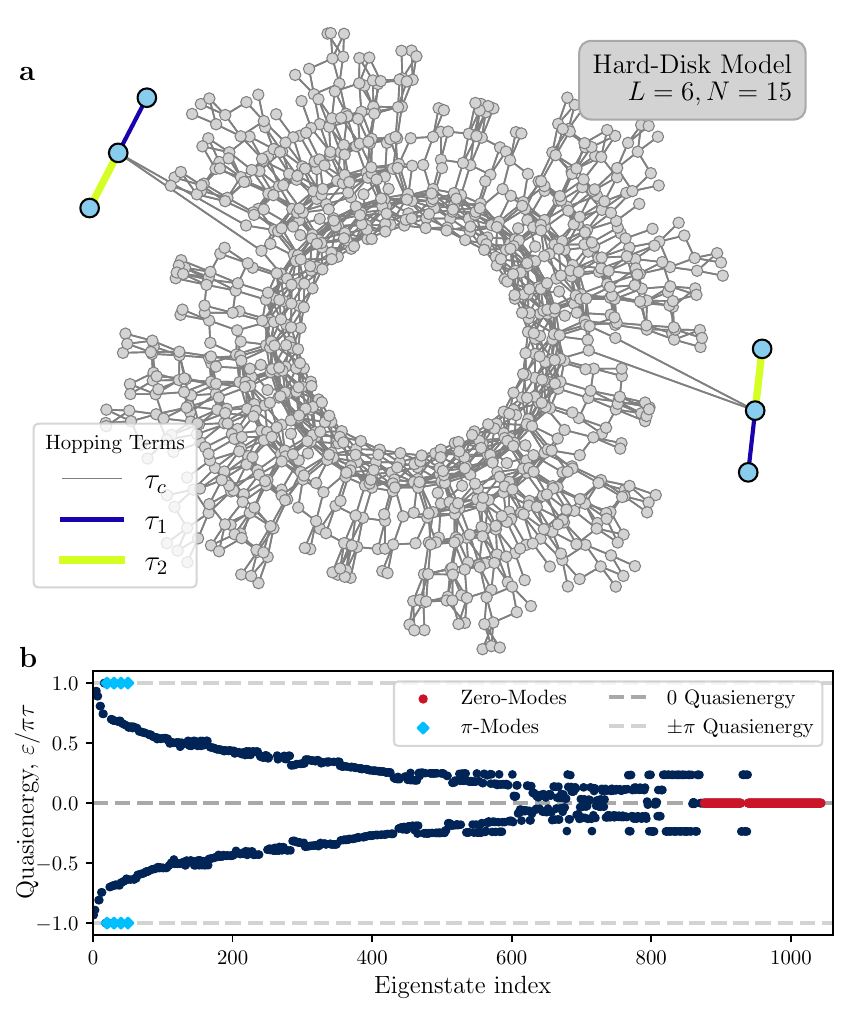}
    {\phantomsubcaption\label{sfig:HD_SSH}}
    {\phantomsubcaption\label{sfig:HD_TC}}
    \caption{\textbf{Engineered many-body cages.} \textbf{a} Many-body state graph of the QHD, with $L=6, N=15$. Examples of grafted three-site trees are highlighted.
    \textbf{b} Quasienergy spectrum of the engineered $\varepsilon=\pi$ edge modes.}
    \label{fig:DTC}
\end{figure}

To build intuition, consider a tree subgraph in the many-body state graph of the QHD, see Fig.~\ref{sfig:HD_SSH}. Under the HV drive, this tree subgraph maps onto a tight-binding model with alternating hopping elements in a manner analogous to what one would have in the Su–Schrieffer–Heeger (SSH) model. This opens up the possibility of engineering MBCs with topological properties inherited from the underlying tree structure.

As noted above, we observe that the MBCs of the QHD model arise form tree-grafting~\cite{ben2025many}, and include three-site tree structures, seen in Fig.~\ref{sfig:HD_SSH}. Remarkably, such trees support a zero-eigenvalue state localised solely on the two end sites with vanishing occupation on the central site. In the presence of an SSH-type modulation of the hopping amplitudes, these zero modes morph into edge modes akin to those seen in the SSH chain.

\phantomsection \label{sec:DTC}
\textit{Engineering Discrete Time Crystal~---~}
Building on these insights, we next engineer a novel eigenspectrum order in the form of a many-body caged discrete time crystal. The key idea is to extend the palindromic drive by an additional swap operation that flips the two end sites of the zero-energy tree state discussed above.
For the QHD model, this swap operation is implemented as a hopping Hamiltonian for the relevant particle, $H_\mathrm{swap} = J_\mathrm{swap} \sum_{\langle\langle l,m\rangle\rangle} P_l \left( a_l^\dagger a_m + \mathrm{h.c.} \right) P_m$, where $\langle\langle l,m\rangle\rangle$ denotes next-to-nearest-neighbour pairs.

The final engineered palindromic drive for one period then becomes $U^* = U_\mathrm{swap} \, U_{HV} \, U_\mathrm{swap}$. 
This construction preserves the overall palindromic symmetry while introducing a $\pi$-phase kick to the zero-energy tree states with $U_\mathrm{swap}^2 = \exp(-i H_\mathrm{swap} \pi/J_\mathrm{swap})$ implementing a full flip operation on the two end sites of the tree.

In Fig.~\ref{sfig:HD_TC}, we display the corresponding quasienergy spectrum of the engineered drive. This engineered palindromic drive yields a MBC with quasienergy at $\epsilon = \pi/\tau$, the hallmark signature of a discrete time crystal. Concretely, this implies that there exist initial conditions where a time-evolved state returns to its initial form only after two periods, thereby breaking the discrete time-translational symmetry of the Floquet dynamics.

The many-body caged discrete time crystal realized here differs fundamentally from conventional discrete time crystals in MBL systems. The spatiotemporal order of the latter is protected by quenched disorder and interactions. In contrast, our caged time crystal owes its existence to \textit{geometric constraints} and \textit{quantum interference} imposed by the local hard-disk constraints and the palindromic drive symmetry, and thus appears in a clean, disorder-free systems through careful engineering of the many-body state graph structure. 

\phantomsection \label{sec:disc}
\textit{Discussion \& outlook~---~}
Generically, driving often destabilises many-body systems~\cite{lazarides2015fate,ponte2015many}, pushing them towards a structureless `infinite-temperature' ensemble~\cite{mori2018thermalization, abanin2017rigorous}.
Floquet MBCs offer a new path to evade such Floquet thermalisation, for appropriately tailored drives. 

Our palindromic construction provides a concrete implementation of Floquet MBCs, with broad applicability. The symmetry-preserving mechanism applies to any quantum circuit structure where chiral symmetry plays a role, making it directly relevant for experimental implementations of MBCs across diverse quantum computing or digital quantum simulation platforms.

Looking forward, several research directions emerge naturally from our work.
The single-particle topological engineering demonstrated here can be straightforwardly extended to pattern further interesting structures. One could imagine engineering complex hopping amplitudes on MBCs with circular structures, which would effectively generate artificial gauge fields on the many-body state graph. Such gauge fields could induce other topological properties in the MBC spectrum potentially yielding novel eigenspectrum order.

Further, our work suggests a broader paradigm: lifting the entire Floquet engineering toolbox explored extensively in real-space particle systems to the many-body state (i.e., Fock) space. Just as Floquet engineering has enabled the creation of topological insulators, anomalous edge states, and time crystals in real-space lattices, similar techniques can now be applied to engineer structured eigenstates in the exponentially large Hilbert space. 

\begin{acknowledgments}
\textit{Acknowledgments -- }
This work was in part supported by the Deutsche Forschungsgemeinschaft under grants FOR 5522 (project-id 499180199) and the cluster of excellence ct.qmat (EXC 2147, project-id 390858490).
This project has received funding from the European Research Council (ERC) under the European Union’s Horizon 2020 research and innovation programme (grant agreement No. 853443).
The authors gratefully acknowledge the resources on the LiCCA HPC cluster of the University of Augsburg, co-funded by the Deutsche Forschungsgemeinschaft (DFG, German Research Foundation) – Project-ID 499211671.

\end{acknowledgments}

\textit{Data availability —} The data to generate all figures in this letter is available in Zenodo \citep{data_repo}

\appendix

\section{End Matter}

\phantomsection \label{app:MBC}
\textit{Many-Body Cages~---~}
Tree grafting is one of the mechanisms which give rise to MBCs in sparse state graphs, often observed in constrained systems~\citep{ben2025many}.
In such sparse state graphs, compact localised states (CLS) emerge on finite trees which are sparsely connected to the state graph. This is captured through the tree-grafting mechanism~\cite{ben2025many, golinelli2003statistics, bauer2001random}, wherein eigenstates of such dangling trees, which have vanishing weight on the connecting node, are also eigenstates of the complete graph with support on all other nodes set to zero, thereby yielding a MBC. 

When multiple dangling trees of the same type are present in the state graph, degenerate flat bands emerge in the many-body spectrum, due to the multiple same-energy CLS that arise in each grafted tree independently. In homogeneous systems, i.e. those where all hopping terms are identical, these many-body flat bands occur at generic energies corresponding to small tree eigenvalues~\cite{ben2025many}.

\vspace{0.8em}
\phantomsection \label{app:HD}
\textit{Quantum Hard-Disk Model~---~}
We define the autocorrelation function for the QHD model as follows:
\begin{equation} \label{eq:autocorr}
    C(t) = \frac{\frac{1}{L^2} \sum_{i} \expval{\qty(2 \hat{n}_i(t) - 1)\qty(2 \hat{n}_i(0) - 1)} - C^*}{1-C^*},
\end{equation}
where $C^* = \left(2\rho - 1\right)^2$ ensures that $C(t) \rightarrow 0$ if no long-time memory exists. 

\begin{figure}[!hbt]
    \centering
    \includegraphics[width=1.0\columnwidth]{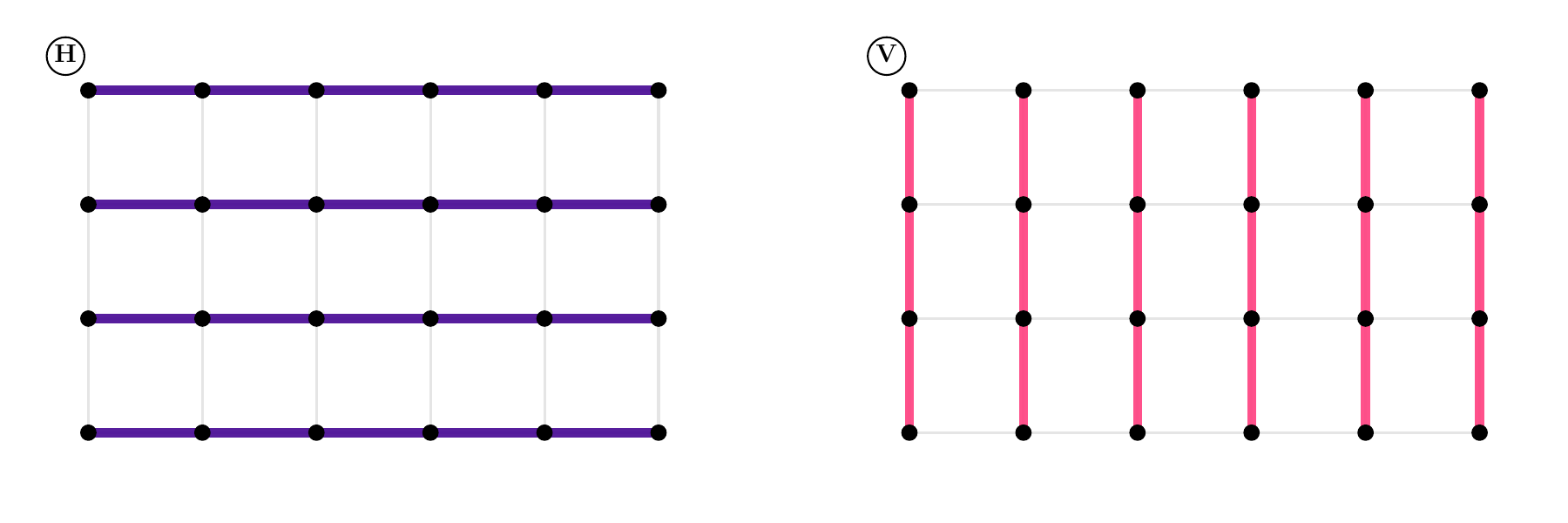}
    \caption{QHD horizontal-vertical drive protocol.}
    \label{fig:drives}
\end{figure}

\vspace{0.4em}
\phantomsection \label{app:HV}
\textit{Horizontal-Vertical Drive~---~}
The Horizontal-Vertical drive protocol (Fig.~\ref{fig:drives}) is a simple, minimal, protocol which only affects the available hopping terms on the QHD model. 
The projectors enforcing the hard-core constraints remain unaffected. Thus, only the allowed directions of hopping depend on time. 
By using a stepped drive which turns on/off horizontal/vertical hopping on the square lattice, we construct a chiral symmetry preserving drive that hosts MBCs at multiple energies. 

The generator for each layer is constructed using the QHD Hamiltonian (Eq.~\ref{eq:QHD}). We thus have
\begin{equation}
    \begin{split}
        H_V & = J\sum_{\left<\vb{r}, \vb{r} + \vu{y} \right>} P_{\vb{r}} \left( a_{\vb{r}}^\dagger a_{\vb{r} + \vu{y}} + a_{\vb{r} + \vu{y}}^\dagger a_{\vb{r}} \right) P_{\vb{r} + \vu{y}},\\
        H_H & = J\sum_{\left<\vb{r}, \vb{r} + \vu{x} \right>} P_{\vb{r}} \left( a_{\vb{r}}^\dagger a_{\vb{r} + \vu{x}} + a_{\vb{r} + \vu{x}}^\dagger a_{\vb{r}} \right) P_{\vb{r} + \vu{x}},
    \end{split}
\end{equation}

\vspace{0.8em}
\phantomsection \label{app:EA_eng}
\textit{Order-Parameter Engineering~---~}
For many-body caged systems, a band-overlap order parameter quantifies the localisation properties within each flat band~\cite{ben2025many}, 
\begin{equation}
    \qea(\ell, \varepsilon) = \sum_{n|E_n = \varepsilon} \abs{\braket{\phi_\ell}{E_n}}^2.
\end{equation}
This order parameter reveals a hierarchy of localised states.
Since Floquet engineering directly modifies many-body cages and affects their CLS, the band-overlap curve allows us to identify how many-body cages are modified by the applied drive, without initially looking for the exact structures on the state graph. 
Thus through the order parameter curve, we can see the effect of the drive on the whole system and the cages within it, and not only specific states.

\begin{figure}[!hbt]
    \centering
    \includegraphics[width=1.0\columnwidth]{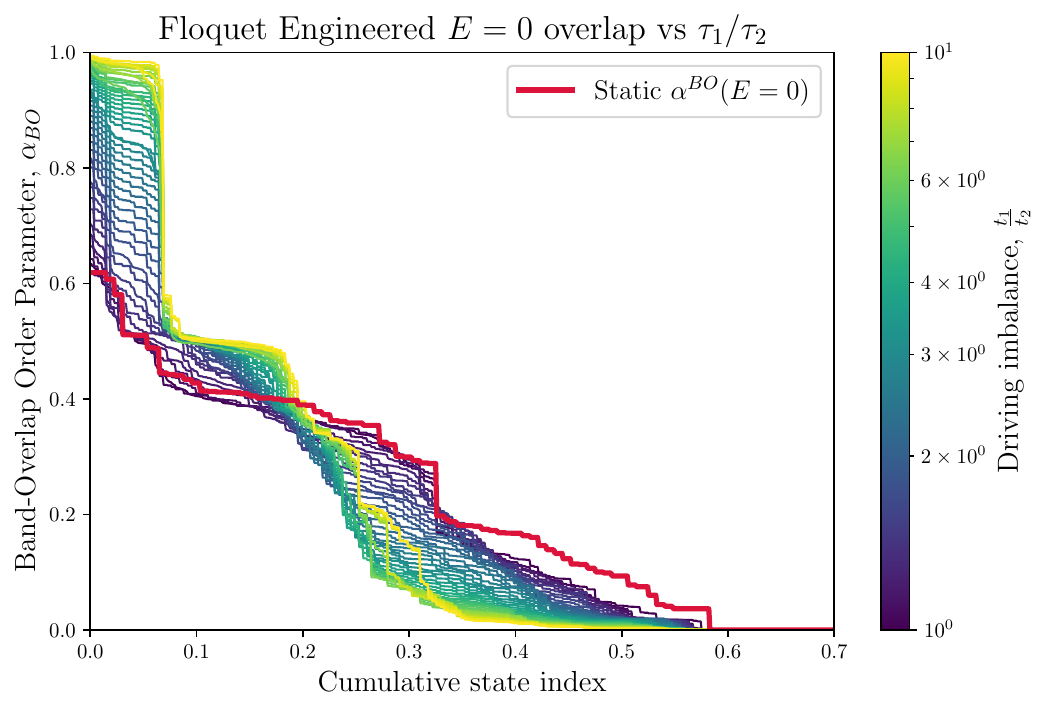}
    \caption{\textbf{Engineered Order-Parameter.} The band-overlap order parameter of the $6\times6$ QHD model with 15 particles, driven using the HV drive. As the driving imbalance, $\tau_1/\tau_2$, is increased, the hierarchy of localised states orders into two steps at approximately $\qea \approx 1.0, 0.5$. The red curve represents the $\qea$ curve for that static system.}
    \label{fig:EAf}
\end{figure}

The exact effect on the $\qea$ curve depends on the system, cage types affected, and the drive used. We demonstrate this using the HV drive with imbalanced durations on the $6\times 6$ QHD model (Fig.~\ref{fig:EAf}). As the driving imbalance $\tau_1/\tau_2$ increases, the $\qea$ curve evolves and two distinct steps, at $\qea = 1.0, 0.5$ appear. The step at $\qea=1.0$ corresponds the the creation of edge states due to the the modulation, which are strongly localised on a single node of the state graph. The static system notably does not include these features. 

\bibliography{references}

\pagebreak                
\onecolumngrid

\appendix
\section*{Supplementary Material: \\ Floquet Many-Body Cages}

\title{Supplementary Material: \\ Floquet Many-Body Cages}

\maketitle

\author{Tom Ben-Ami}
\affiliation{Max-Planck-Institut f\"{u}r Physik komplexer Systeme, N\"{o}thnitzer Stra\ss e 38, Dresden 01187, Germany}
\affiliation{Theoretical Physics III, Center for Electronic Correlations and Magnetism, Institute of Physics, University of Augsburg, D-86135 Augsburg, Germany}

\author{Roderich Moessner}
\affiliation{Max-Planck-Institut f\"{u}r Physik komplexer Systeme, N\"{o}thnitzer Stra\ss e 38, Dresden 01187, Germany}

\author{Markus Heyl}
\affiliation{Theoretical Physics III, Center for Electronic Correlations and Magnetism, Institute of Physics, University of Augsburg, D-86135 Augsburg, Germany}
\affiliation{Centre for Advanced Analytics and Predictive Sciences (CAAPS), University of Augsburg, Universitätsstr. 12a, 86159 Augsburg, Germany}

\section{Symmetric Anomalous Floquet Anderson Insulator Drive}

In addition to the horizontal-vertical drive introduced, we look at a symmetric version of the anomalous Floquet Anderson drive (AFAI)~\cite{rudner2013anomalous} for the quantum hard-disk model.

\begin{figure}[!hbt]
    \centering
    \includegraphics[width=1.0\columnwidth]{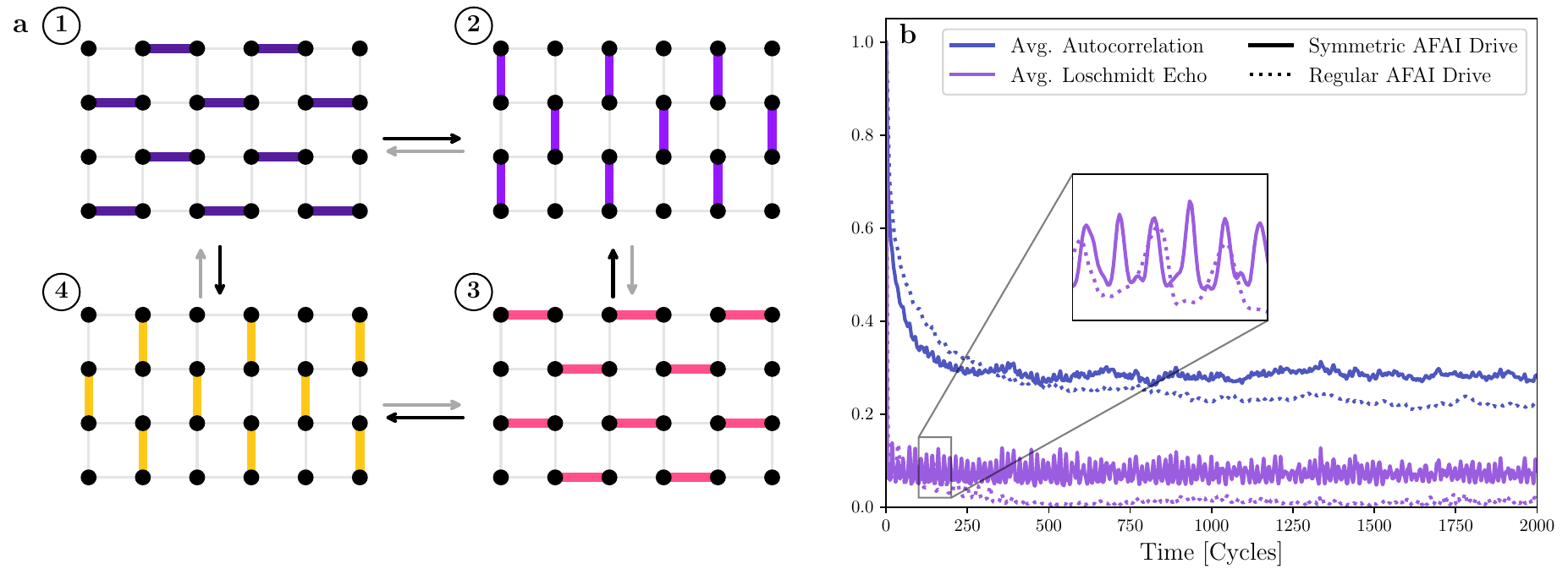}
    {\phantomsubcaption\label{sfig:AFAI}}
    {\phantomsubcaption\label{sfig:AFAI_LE}}
    \caption{\textbf{AFAI Driven QHD.} \textbf{a} Symmetric AFAI drive steps. \textbf{b} Long-time memory under the symmetric AFAI drive. The inset shows many-body Rabi oscillations of the averaged Loschmidt Echo.}
    \label{fig:AFAI}
\end{figure}

The AFAI drive (Fig.~\ref{sfig:AFAI}) consists of modulating the hopping terms on a square lattice in a chiral pattern around plaquettes:
\begin{enumerate}
    \item $U_{1}$ - Odd horizontal hopping
    \item $U_{2}$ - Odd vertical hopping
    \item $U_{3}$ - Even horizontal hopping
    \item $U_{4}$ - Even vertical hopping
\end{enumerate}
The standard AFAI drive follows:
\begin{equation}
    U_F = U_{1}(\tau) U_{2}(\tau) U_{3}(\tau) U_{4}(\tau),
\end{equation}
while we symmetrise this in time, creating a chiral symmetry peserving palindromic drive:
\begin{equation}
    U_F = U_{1}(\frac{\tau}{2}) U_{2}(\frac{\tau}{2}) U_{3}(\frac{\tau}{2}) U_{4}(\tau) U_{3}(\frac{\tau}{2}) U_{2}(\frac{\tau}{2}) U_{1}(\frac{\tau}{2}).
\end{equation}

We find that the symmetric AFAI drive preserves long-time memory, while the standard AFAI drive, which breaks the MBCs in the hard-disk model, shows no long-time memory. In the inset, we find similar Rabi oscillations as those found in the horizontal-vertical drive. 

\section{Engineered Cages - Toy Model}

Since the compact localised states which arise from many-body cages are decoupled from the bulk of the state graph through destructive interference, their properties are independent to its specifics far away. 

\begin{figure}[!hbt]
    \centering
    \includegraphics[width=1.0\linewidth]{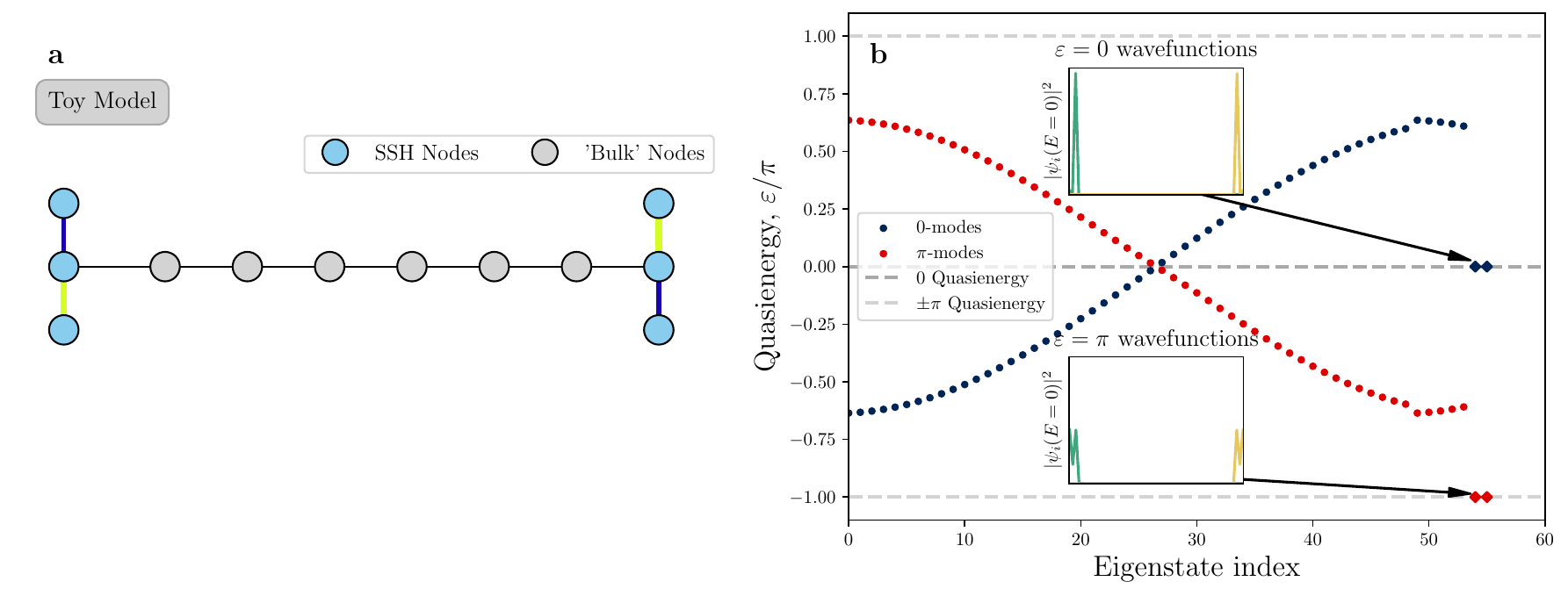}\;
    {\phantomsubcaption\label{sfig:ToyM}}
    {\phantomsubcaption\label{sfig:ToySpec}}
    \caption{\textbf{Toy Model State Graph \& Spectrum.} \textbf{a} - A toy-model recreating two grafted trees connected via a long 1D chain (The `bulk'). Additionally, it only hosts two many-body cages, corresponding to the two grafted trees at each end.
    \textbf{b} - Quasienergy spectrum of the toy model with both zero-modes (Dark Blue) and $\pi-$modes (Red). A `bulk chain' of $L=50$ is used here.}
    \label{fig:toy}
\end{figure}

Therefore, we construct a toy model comprising two grafted 3-site trees, akin to those identified in the quantum hard-disk model in Fig.~4a. We connect them at their central nodes to a long chain (Fig.~\ref{sfig:ToyM}), which replaces the `bulk' of the quantum hard-disk model. This setup hosts exactly two many-body cages, enabling us to directly manipulate and isolate them from the rest of the zero-energy CLS, hosted on cages with different geometries.
Applying the an analogue of the HV Floquet drive protocol, the quasienergy spectrum includes two zero-energy edge modes (see Fig. \ref{sfig:ToySpec}), each localised at the edge of its respective tree.

We additionally introduce a swap operator analogous to the one introduced for the quantum hard-disk model, which in turn elevates the zero edge-modes to $\pi-$modes, as shown in Fig.~\ref{sfig:ToySpec}.


\end{document}